\documentstyle[11pt]{article}

\begin{document}
% \begin{titlepage}

\title{Space-time dimension, Euclidean action and signature change
         \thanks{Talk given at the $3^{\rm rd}$ Alexander Friedmann
                international seminar on gravitation and
                cosmology (St. Petersburg, July 4 -- 12, 1995).
                 }}

\author{Franz Embacher\\
        Institut f\"ur Theoretische Physik\\
        Universit\"at Wien\\
        Boltzmanngasse 5\\
        A-1090 Wien\\
        \\
        E-mail: fe@pap.univie.ac.at\\
       \\
        UWThPh-1995-28\\
        gr-qc/9507041
        }
\date{}

\maketitle

\begin{abstract}
This talk is devoted to the problem how to compute relative
nucleation probabilities of configurations with different topology and
dimension in quantum cosmology. Assuming the semiclassical approximation,
the usual formula for the nucleation probability induced by the
no-boundary wave function
is $P_{NB}\approx\exp(-I)$, where $I$ is the
Euclidean action, evaluated at a solution of the effective
Euclidean field equations. In the simplest case, these are just
Einstein's field equations with a cosmological constant $\Lambda$.
Relative probabilities of different configurations are usually compared
at equal values of $\Lambda$. If $\Lambda$ is an effective
vacuum energy density arising from, say, a massive scalar field
$\phi$ (i.e. $\Lambda\sim \phi^2$), one thus compares probabilities at
equal values of this
field. When configurations with different dimensions are admitted
(the $n$-dimensional gravitational
constant being subject to a rather mild restriction), as e.g.
${\bf S}^n$ for any $n$, this procedure leads to the prediction that
the space-time dimension tends to be as large as possible,
$n\rightarrow\infty$. In this contribution, I would like to
propose an alternative scheme, namely to compare the probabilities
$P_{NB}\approx\exp(-I)$ at equal values of the {\it energy} $E$,
instead of the {\it energy density} $\Lambda$.
As a result, the space-time dimension settles at $n=4$.
Attempts to predict the topology of the spacelike slices lead to
the candidates
${\bf S}^3$ and ${\bf S}^1\times {\bf S}^2$.
Since the ''process'' of nucleation (possibly connected with decoherence)
is not well known in detail, we expect that either {\it both}
configurations may be realized with roughly equal probability, or
the {\it latter} one is favoured.
Finally, we comment on the analogous situation based on the tunneling
wave function.
\medskip

% PACS-numbers:
\end{abstract}

% \end{titlepage}

\section{Euclidean quantum cosmology, signature change
and nucleation probabitilies}

The no-boundary wave function of the universe for a model whose variables
are the metric $g_{\mu\nu}$ and some matter fields (denoted as
$\Phi$ and $\phi$) has as its arguments the values $(h_{ij},\Phi,\phi)$
of the spatial metric and the matter fields, evaluated at some spacelike
hypersurface $\Sigma$. It is symbolically given by a path
integral
\cite{HartleHawking}
\begin{equation}
\psi_{NB}[h_{ij},\Phi |_\Sigma,\phi |_\Sigma] =
\int {\cal D}g\, {\cal D}\Phi\, {\cal D}\phi\, e^{-I}
\label{1.1}
\end{equation}
over compact Euclidean metrics (i.e. Riemannian metrics:
signature $++\dots+$) and according Euclidean matter
configurations. A standard procedure to approximate this object
is to replace the path integral over generic configurations
by a sum over configurations that solve the effective Euclidean field
equations
\cite{111}.
Here, by ''effective'', we refer to a division of
the matter variables into two groups, denoted by $\Phi$ and
$\phi$, such that the field equations become
\begin{equation}
R_{\mu\nu} - \,\frac{1}{2}\,g_{\mu\nu} R = - \Lambda \,g_{\mu\nu}
+ 8 \pi G_n\, T_{\mu\nu}(\Phi)\,,
\label{1.2}
\end{equation}
together with the field equation for $\Phi$.
$\Lambda\equiv \Lambda(\phi)$ is an effective cosmological constant.
In case of $\phi$ being a minimally coupled scalar
field with potential $V(\phi)$, it becomes
$\Lambda(\phi)=8\pi G_n V(\phi)$ and defines a regime in which the
approximation $\phi\approx const$ is imposed (cf. Ref.
\cite{Hawking2}).
$G_n$ is the $n$-dimensional
gravitational constant.
\medskip

The division of the matter variables into $\Phi$
and $\phi$ is somewhat arbitrary and corresponds to the concrete
goals one has in mind. For fixed $\Lambda$, the wave function
behaves exponential for ''small'' spatial geometries (small spatial
volumes) and oscillatory for ''large'' spatial geometries. In
between these regimes the universe is thought to become ''real''
or classical. $\psi_{NB}$ develops into a WKB-type
wave function peaked around a family of classical Lorentzian
(i.e. Pseudo-Riemannian) time evolutions (signature $-++\dots+$)
\cite{Halliwell3}.
This is usually refered to as ''nucleation'' of the universe.
However, there must be an additional step that explains why
only one member of this family is observed. This may either
be viewed as a property emerging dynamically (''decoherence'';
see e.g. Refs. \cite{GellMannHartleHartle}) or
(at least to some extent) be achieved by means of the physical
interpretation of the wave function ({\it \`a la} the observer
herself is part of just one classical history).
\medskip

In the semiclassical approximation any nucleation
scenario may be encoded in terms of a ''real tunneling configuration''
\cite{GibbonsHartle}.
By this we mean a manifold ${\cal M}_{\rm sig\,\,ch}$, divided into two parts
${\cal M}_{\rm Eucl}$ and ${\cal M}_{\rm Lor}$ by a hypersurface
$\Sigma$ such that the Euclidean field equations for
$(g_{\mu\nu},\Phi)$ with cosmological constant
$\Lambda(\phi)$ are satisfied on
${\cal M}_{\rm Eucl}$ (which is assumed to be
compact), and the according Lorentzial (physical) field equations
are satisfied on ${\cal M}_{\rm Lor}$. The junction conditions
at the hypersurface $\Sigma$ (at which the metric signature
changes discontinuously) are that the extrinsic curvature $K_{ij}$
as well as the (affine) time derivative $\partial_t \Phi$ of the
matter fields vanish and that $\Sigma$ is spacelike
with respect to the Lorentzian part (see Fig. 1).
The hybrid configuration as a whole
thus satisfies the Einstein equations with matter (which are
{\it \`a priorily} defined without reference to the metric signature)
in a distibutional sense
\cite{222}.
Likewise, it can be regarded as a stationary
point of the action if the latter is defined by
using an imaginary time
variable in the Euclidean domain (Wick rotation). This situation is
sometimes denoted as ''strong'' signature change
\cite{FE4}.
The simplest example occurs if there are no fields $\Phi$,
${\cal M}_{\rm Eucl}$ being half of ${\bf S}^n$ with radius $a$
($2 \Lambda a^2 = (n-1)(n-2)$), and
${\cal M}_{\rm Lor}$ the corresponding half of $n$-dimensional
de Sitter space, joined together along the equator
$\Sigma={\bf S}^{n-1}$ of ${\bf S}^n$
(see Ref. \cite{Vilenkin1}).
\medskip

The information stored in such a configuration is threefold: At the
initial hypersurface $\Sigma$ the universe starts its classical
evolution with some initial values $(h_{ij},\Phi)$, such that
$K_{ij}=\partial_t\Phi=0$. The subsequent classical period
is given by the manifold ${\cal M}_{\rm Lor}$ and the field configuration
on it. The probability for this scenario (relative to others of the
same type) is given by
\begin{equation}
P_{NB} = |\psi_{NB}|^2 \approx e^{- 2 I_{\rm Eucl}}
\label{1.3}
\end{equation}
where $I_{\rm Eucl}$ is the Euclidean action, evaluated over the
Euclidean part
${\cal M}_{\rm Eucl}$. Hence, $P_{NB}\equiv P_{NB}(\phi)$
may in principle distinguish between different candidate
nucleation configurations. Since
${\cal M}_{\rm Eucl}$ is half of a compact configuration
${\cal M}$ admitting a reflection symmetry (in the above
example
this is just the whole of ${\bf S}^n$), one may write as well
$P_{NB}\approx \exp(-I)$, with $I$ the Euclidean action evaluated
over $\cal M$.
\medskip

The standard pocedure to analyze a particular model within the
range of validity of the approximations imposed consists of
$(i)$ considering all solutions to the Euclidean field equations
which are regular and admit a hypersurface $\Sigma$ along
which they may be joined to their Lorentzian counterparts;
$(ii)$ minimizing $I$ (i.e. maximizing $P_{NB}$) at constant $\Lambda$
and $(iii)$ comparing the maximized values of $P(\phi)$ at different
$\phi$.
The interpretation of the third step depends on the details
and the goals connected with the particular
model. In the case of a massive scalar field $\phi$,
the prediction is that the nucleation value (i.e. initial value)
of $\phi$ tends to be as small as possible (thus getting thrown out
of the range of applicability of the approximation $\phi\approx const$).
Nevertheless, in the literature configurations with different topology
(as e.g. ${\bf S}^4$ and ${\bf S}^2\times {\bf S}^2$)
are frequently compared by means of $P_{NB}$
at equal (finite) values of $\phi$ (respectively $\Lambda$;
see e.g. Refs. \cite{GibbonsHartle} and
\cite{BoussoHawking}).
A somewhat different sort of models emerges if
small scale topological fluctuations
(wormholes) are admitted ($\phi$ representing the
''wormhole parameters''). Then the procedure outlined above leads to
Coleman's argument that $\Lambda$ is infinitely peaked at $0$
\cite{333}.
In the following we are mainly interested in
the question of ''nucleation''. Moreover, we will leave
apart the matter fields denoted by
$\Phi$, so the field equations reduce to Einstein's equations
with a (positive) effective cosmological constant $\Lambda(\phi)$.
In the case of a minimally coupled massive scalar field,
$\Lambda(\phi) = 4\pi G_n m^2 \phi^2$.
\medskip

\section{Nucleation energy}
\setcounter{equation}{0}

The effective cosmological constant plays the role of a vacuum
{\it energy density}. To be more precise,
at the nucleation hypersurface $\Sigma$, the non-gravitational
energy density is given by the expression $\Lambda/(8\pi G_n)$.
Hence (omitting any further field $\Phi$), the corresponding
amount of {\it energy} with which the universe is born, is given by
\begin{equation}
E = \int_\Sigma d^{n-1}x\, \sqrt{h} \,\frac{\Lambda}{8\pi G_n}
= {\cal V}_\Sigma \,\,\frac{\Lambda}{8\pi G_n},
\label{2.1}
\end{equation}
where ${\cal V}_\Sigma$ is the volume of $\Sigma$ as an $(n-1)$-manifold.
However, note that the total energy contained in a closed universe
is in some sense identical to zero (since the gravitational field
carries precisely the energy $-E$).
\medskip

Now suppose we are given several possible nucleation
configurations at equal values of $\Lambda$, each one
sloppily denoted by $({\cal M},\Sigma)$. Minimizing
$I$ (maximizing $P_{NB}$) at constant $\Lambda$ (and
constant $n$) corresponds to
the question for the most probable classical
initial configuration, given that the energy density at nucleation is
$\Lambda/(8\pi G_n)$.
Since for (low dimensional) products of spheres the radii
appearing are of the order $a\sim \Lambda^{-1/2}$,
this is somewhat related to
the question for the most probable nucleation
configuration at a given ''size'' of the initial hypersurface
$\Sigma$. Thereby the underlying idea is that the wave function
describes an ensemble of universes that ''probe'' for
competing configurations at equal values of the energy density
or of the size. This idea is related to the use of the
''configuration representation'' $\psi_{NB}[h_{ij},\dots]$.
Once a configuration has beeen singled out (e.g. by
overwhelming probability or by some process of decoherence),
we may naively think about the universe being created at
some initial energy density or at some initial size.
\medskip

It is however conceivable that some sort of energy $E$ is included in the
variables, as $\psi_{NB}[E,\dots]$, such that for the
nucleation configuration $E$ coincides with the quantity defined
above. This would amount to a different question:
Which of several competing configurations is most probable,
when the comparison is carries out at equal values of $E$?
When some configuration is realized, we may naively think
about the universe being created at some initial energy.
In a formal sense, this would relate the emergence of classical
time with its conjugate quantity, energy.
\medskip

One would expect both questions to give the same answers,
at least as far as large scale variables such as topology and
dimension are concerned. However, this is not the case, the formal
reason being that the relation (\ref{2.1}) between $\Lambda$ and
$E$ contains the volume ${\cal V}_\Sigma$ as well as the
dimension $n$.
\medskip

Since we do not know much about the nature of the underling
structures (we do not even know to what extent these
structures are physical ''processes'' that depend on the
particular model, or fundamental issues related to the interpretation of
wave function of the universe), it is worth
investigating the structure of the relative
probabilities for competing configurations
if the energy $E$ is kept fixed, instead of the cosmological constant.
\medskip

{}From now on we admit configurations at arbitrary topology and
dimension, hence we consider a ''multiple-dimensional''
model, rather than just a ''multi-dimensional'' one.
(Different approaches to the problem of space-time dimension
in Euclidean quantum cosmology may be found in
Ref. \cite{444}).
The $n$-dimensional gravitational constant is written as
\begin{equation}
G_n = \left(\frac{\kappa_n}{m_P}\right)^{n-2}
\label{2.2}
\end{equation}
where we just know $\kappa_4 = 1$. In the main part of my talk I
will use $\kappa_n\approx 1$ for all $n$, although this condition
may be relaxed without changing much of the results.
\medskip

\section{Probabitities at equal energy}
\setcounter{equation}{0}

Minimizing the Euclidean action at constant $\Lambda$ leads to
the problem of arbitrarily large dimensions. Inserting
round spheres ${\cal M} = {\bf S}^n$ with radius $a$, the
Euclidean Einstein equations
imply $a^2 =(n-1)(n-2)/(2\Lambda)$, and for large $n$
\begin{equation}
I \sim - \left(\frac{n}{\Lambda}\right)^{n/2}\,.
\label{3.1}
\end{equation}
Hence, there is no finite minimizing dimension $n$.
\medskip

In order to test the proposal formulated above, I would like to
admit a larger set of configurations. Consider arbitrary products
of round spheres
\begin{equation}
{\cal M} = {\bf S}^{n_1} \times {\bf S}^{n_2} \times \dots
\times \widetilde{  {\bf S}^{n_A}  } \times\dots
\times {\bf S}^{n_m}
\label{3.2}
\end{equation}
with radii $(a_1,\dots,a_m)$ and total dimension
$n =\sum_{B=1}^m n_B$. In the $A$-th sphere (denoted by a
$\widetilde{{}\,\,\,}$) the change of signature occurs, i.e.
this sphere is joined along its equator to half of the
$n_A$-dimensional de Sitter space, whereas the other spheres
remain unaffected. (In other words, the Lorentzian time
coordinate emerges from an angular coordinate on the $A$-th
factor sphere by a Wick rotation). The nucleation hypersurface $\Sigma$
is thus the product of ${\bf S}^{n_A-1}$ with all the remaining
spheres ${\bf S}^{n_B}$. The Euclidean Einstein equations
reduce to $2\Lambda a_B^2 = (n_B-1)(n-2)$, and hence
require
all $n_B>1$ and $n\geq 3$. As a consequence, all the radii $a_B$ are
completely determined by $\Lambda$ and the dimensions $n_B$.
The Euclidean action has the form
\begin{equation}
I = -\, \frac{1}{16 \pi G_n}
\int_{\cal M} d^n x\,\sqrt{g}\,\left(R-2\Lambda\right)
\label{3.3}
\end{equation}
(note that no boundary term is needed here). This can be
evaluated on the set of solutions specified above. Our
proposal consists of eliminating $\Lambda$ in terms of $E$,
which yields, after some tedious manipulations,
\begin{equation}
I =
- \left(\frac{8\pi} F \right)^{1/(n-3)}
\left(  \frac{2\, \kappa_n}{n-2}\,\frac{E}{m_P}
\right)^{(n-2)/(n-3)} \,,
\label{3.4}
\end{equation}
where
\begin{equation}
 F  =
 \left(
\frac{v_{n_A-1}}{v_{n_A} (n_A-1)^{1/2}} \right)^{n-2}
\prod_{B=1}^m v_{n_B} (n_B-1)^{n_B/2}\,\,,
\label{3.5}
\end{equation}
$v_q$ being the volume of the unit-${\bf S}^q$.
If $n=3$, the energy is fixed by $E=1/(6 G_3)$, and since we expect
$E$ to play the role of a generic quantity, we ignore this case
and set $n\geq 4$. We analyze this model in three steps, the
mathematical details of which have originally been presented
in Ref. \cite{FE5}.
\medskip

{\bf Step 1: Minimize $I$ at $E$ and $n$ fixed.} This amounts to
minimize $F$, and we do not have to know the $\kappa_n$ during
this first step. The analysis for small $n$ may be carried out by
explicitly by computing the quantity $F$.
At any $n$, we find a Euclidean configuration ${\cal K}_n$
that minimizes the action $I$. The first few of these configurations
are
\begin{eqnarray}
{\cal K}_4 &=& {\bf S}^2\times\widetilde{{\bf S}^2}
\label{lll4}\\
{\cal K}_5 &=& {\bf S}^2\times\widetilde{{\bf S}^3}
\label{lll5}\\
{\cal K}_6 &=& {\bf S}^2\times\widetilde{{\bf S}^4}
\label{lll6}\\
{\cal K}_7 &=&
{\bf S}^2\times{\bf S}^2\times\widetilde{{\bf S}^3}\,.
\label{lll7}
\end{eqnarray}
It is quite surprising that at $n=4$ the favoured configuration
is not the round ${\bf S}^4$. The nucleation hypersuface associated with
${\cal K}_4$ is $\Sigma={\bf S}^1\times {\bf S}^2$.
For large $n$ one obtains that ${\cal K}_n$ is a product of
a bunch of
two-spheres with ${\bf S}^p$, where $p\approx 1.277 \sqrt{n}+1$.
\medskip

Hence, the favoured topology of the spatial sections in $n=4$
is ${\bf S}^1\times {\bf S}^2$, which implies (as far as it is
actually realized) that the universe is
of Kantowski-Sachs type. This topology may as well be interpreted as
representing an ${\bf S}^3$ with a (primordial) black hole
\cite{BoussoHawking}.
\medskip

{\bf Step 2: Minimize $I({\cal K}_n)$ at fixed $E$.} For large
$n$ we find
\begin{equation}
I({\cal K}_n) = -\sqrt{2} \left(\frac{\kappa_n}{n-2}
\right)^{(n-2)/(n-3)} \frac{E}{m_P}\left( 1 + O(\frac{1}{n})\right)
\sim - \,\frac{1}{n}\,\frac{E}{m_P}
\label{3.6}
\end{equation}
where the $\sim$ sign is for $\kappa_n\approx 1$. This may be
relaxed to the condition $\kappa_n / n \rightarrow 0$ and possibly
some monotonicity requirement without changing much. As a consequence,
with each energy $E$ we may associate a
well-defined dimension $n$ minimizing (\ref{3.6}).
Asymptotically one finds, $n\sim \ln(m_P/E)$. Hence,
large $E$ corresponds to small $n$. For large $E$, the exponent
structure in (\ref{3.4}) implies that the minimizing dimension is
$n=4$. At some value $E=E_4$, defined by the equality
$I({\cal K}_4) = I({\cal K}_5)$, the minimizing dimension becomes
$n=5$. Defining by analogous equality of adjacent probabilities
a sequence of energy levels $E_n$, we find $n$ to be the
minimizing dimension in the interval $E_n<E<E_{n-1}$. The
first two value (for $\kappa_5=\kappa_6=1$)
are $E_4 \approx 0.287 m_P$ and
$E_5\approx 0.143 m_P$.
For large $n$ we find $E_n\sim m_P \exp(-n)$. Fig. 2
shows the minimizing dimension $n$ as a function of $E$
(the scale being $m_P=1$).
The regime of small energies can be thought of as representing
the multiple-dimensional quantum state of the universe.
\medskip

{\bf Step 3: Minimize $I({\cal K}_{\rm minimizing\,\,n})$ with
respect to $E$? } Formally, the action is minimized for
$E\rightarrow\infty$, which implies $n=4$. However, it is not
quite clear what this means physically. One may imagine that
-- in some semiclassical picture -- the universe ''evolves''
from small to large energies. At $E>E_4\approx m_P$, the
minimizing configuration $\Sigma={\bf S}^1\times {\bf S}^2$
may be thought of ''freezing out'' by some decoherence effect.
However, at $E\approx E_4$, several other configurations
will still have comparable probability. Hence we estimate the
relative probabilities with which higher dimensions and
alternative topologies are suppressed. For $E\gg E_4$, we find
\begin{equation}
p_{\rm dim}(E)\equiv \frac{P({\cal K}_5)}{P({\cal K}_4)}\approx
\exp\left(-\frac{2 E^2}{\pi m_P^2}\right) \approx
\left(  \frac{P({\bf S}^4)}{P({\cal K}_4)}
\right)^9\equiv p_{\rm top}(E)^9\,.
\label{3.7}
\end{equation}
Fig. 3 shows these two curves (the scale is $m_P=1$;
for small $E$, the formulae
above are not very accurate, hence the slight
mismatch between (\ref{3.7}) and Fig. 3).
Thus we estimate that at a scale $E_{\rm dim}$ of several $m_P$
the dimensions greater that 4 become suppressed, while at
a larger scale $E_{\rm top} \approx 3 E_{\rm dim}$ the
competing sphere ${\bf S}^4$ becomes suppressed as compared to
${\cal K}_4$.
\medskip

I can imagine three possibilities for a decoherence process to
work: The first one is based on the idea that the relaxation
of the dimension $n=4$ at $E_{\rm dim}$ somehow ''induces''
one of the configurations to ''nucleate''. In this case
$\Sigma={\bf S}^3$ and $\Sigma={\bf S}^1 \times {\bf S}^2$
will be realized at probabilities of the same order of
magnitude. Alternatively, the relaxation of dimension and
topology might be ''decoupled'' from one another and occur
at $E_{\rm dim}$ and $E_{\rm top}$, respectively. (Naively,
''first'' the dimension becomes classical, while the topology
is still quantum). Such a mechanism could result into a
considerable suppression of the isotropic configuration
and predict $\Sigma={\bf S}^1 \times {\bf S}^2$. In both cases,
a scalar field $\phi$ with mass $m\approx 10^{-5} m_P$ (according
to the bound set by the microwave anisotropy) would
nucleate at $\phi\approx m_P^2/m$, which is quite enough
to ensure sufficient inflation of the subsequent classical
time evolution. A third possibility is that nucleation and
decoherence occur at higher energy scales, induced by some
other mechanism. Since there is no charactistic scale above
$E_4$ in our approach, this could mean that we predict
$\phi\rightarrow 0$, and reproduce the usual problems
related with finding the most probable initial scalar field value
for the no-boundary wave function
\cite{555}. This third possibility will throw
us out of the range of our approximations, but the dominance
of $\Sigma={\bf S}^1 \times {\bf S}^2$ might still survive.
\medskip

We should thus study in a conceptually deeper way the relations
between energy, dimensional and topological ''relaxation'',
nucleation and decoherence, emergence of time and the interpretation
of the wave function. For those who do not consider
higher dimensions to be of theoretical relevance at all, there is
still the open question of the
competition between ${\bf S}^4$ and ${\bf S}^2\times {\bf S}^2$.

\section{Tunneling wave function}
\setcounter{equation}{0}

The tunneling wave function of the universe
\cite{Vilenkin1}, \cite{666}
differs from the
no-boundary wave function in the semiclassical Euclidean regime only by
a  sign in the exponent. Thus the nucleation probabilities
may be approximated by $P_T\approx \exp(I)$. Now recall that
the probabilities due to the no-boundary
wave function $P_{NB}\approx\exp(-I)$ can predict
$n=4$ only if the $n$-dimensional gravitational constant
is such that $\kappa_n/n\rightarrow 0$ as $n\rightarrow\infty$
(recall (\ref{2.2}) and (\ref{3.6})).
In the case $\kappa_n/n\rightarrow \infty$, there is a
well-defined dimension $n$ maximizing the action at a given
value of $E$.
Hence, in this case, the tunneling wave function may be invoked.
As opposed to the no-boundary case, small energies correspond
to low dimensions. For sufficiently small $E$
($E<3\sqrt{2}\, \kappa_5^3\, m_P /16$ if some
monotonicity in $n$ is assumed), the favoured dimension
is $n=4$, and the according Euclidean configuration is
${\bf S}^4$, thus $\Sigma={\bf S}^3$. The multiple-dimensional quantum
state of the universe is associated with large values of $E$,
as if the coming-into-existence is related with some energy
minimization.
The Euclidean configurations maximizing the action
at given $E$ and $n$ are, for $n>4$, just
${\bf S}^{n-2} \times \widetilde{{\bf S}^2}$ (as opposed
to ${\cal K}_n$ in the no-boundary case).
Again, the details of the
interpretation would require more knowledge about the nucleation
process.
\medskip

The choice $\kappa_n \sim n$ for large $n$ may be considered
as a limiting case that potentially neutralizes the question for the
space-time dimension.
\\
\\
\\
{\Large {\bf Acknowledgments}}
\medskip

This work was supported by the Austrian Academy of Sciences
in the framework of the ''Austrian Programme for
Advanced Research and Technology''. Also, I would like
to express my thanks to the organizers of this seminar.


\begin{thebibliography}{00}

\bibitem{HartleHawking}
{
 J. B. Hartle and S. W. Hawking,
 {\it Phys. Rev. D} {\bf 28}, 2960 (1983).
}

\bibitem{111}
{
 S. W. Hawking,
{\it Phys. Rev. D} {\bf 18}, 1747 (1978);
 G. W. Gibbons, S. W. Hawking and M. J. Perry,
 {\it Nucl. Phys. B} {\bf 138}, 141 (1978).
 K. Schleich,
 {\it Phys. Rev. D} {\bf 36}, 2342 (1987).
 S. W. Hawking,
 in: S. W. Hawking and W. Israel (eds.),
 {\it General relativity: An Einstein Centenary Survey},
 Cambridge University Press (Cambridge, 1979).
}

\bibitem{Hawking2}
{
 S. W. Hawking,
 {\it Nucl. Phys. B} {\bf 239}, 257 (1984).
}


\bibitem{Halliwell3}
{
 J. J. Halliwell,
 in: S. Coleman {\it et. al.} (eds.),
 {\it Quantum cosmology and baby universes},
 World Scientific (Singapore, 1991), p. 159.
}

\bibitem{GellMannHartleHartle}
{
 M. Gell-Mann and J. B. Hartle,
 in: W. H. Zurek (ed.), {\it Complexity, Entropy and the Physics
 of Information}, Santa Fe Institute Studies in the Sciences
 of Complexity, vol IX (Addison Wesley, 1990).
 J. B. Hartle,
 in: S. Coleman {\it et al} (eds.),
 {\it Quantum cosmology and baby
 universes}, World Scientific (Singapore, 1991).
}


\bibitem{GibbonsHartle}
{
 G. W. Gibbons and J. B. Hartle,
 {\it Phys. Rev. D} {\bf 42}, 2458 (1990).
}

\bibitem{222}
{
 S. A. Hayward,
 {\it Class. Quantum Grav.} {\bf 9}, 1851 (1992);
 {\bf 10}, L7 (1993).
 T. Dereli and R. W. Tucker,
 {\it Class. Quantum Grav.} {\bf 10}, 365 (1993).
 T. Dereli, M. {\"O}nder and R. W. Tucker,
 {\it Class. Quantum Grav.} {\bf 10}, 1425 (1993).
 M. Kossowski and M. Kriele,
 {\it Class. Quantum Grav.} {\bf 10}, 2363 (1993);
 {\bf 10}, 1157 (1993);
 {\it Proc. R. Soc. Lond. A} {\bf 444}, 297 (1994);
 {\bf 446}, 115 (1994).
 M. Kriele and J. Martin,
 {\it preprint} gr-qc/9411063.
}

\bibitem{FE4}
{
 F. Embacher,
 {\it Phys. Rev. D} {\bf 51}, 6764 (1995).
}


\bibitem{Vilenkin1}
{
 A. Vilenkin,
 {\it Phys. Lett. B} {\bf 117}, 25 (1982).
}

\bibitem{BoussoHawking}
{
 R. Bousso and S. W. Hawking,
 {\it preprint} gr-qc/9506047.
}

\bibitem{333}
{
 S. Giddings and A. Strominger,
 {\it Nucl. Phys. B} {\bf 307}, 854 (1988).
 S. Coleman,
 {\it Nucl. Phys. B} {\bf 307}, 867 (1988);
 {\bf 310}, 643 (1988).
}


\bibitem{444}
{
I. V. Volovich,
 {\it Phys. Lett. B} {\bf 219}, 66 (1989).
 M. Gasperini,
 {\it Phys. Lett. B} {\bf 224}, 49 (1989).
 R. C. Myers,
 {\it Nucl. Phys. B} {\bf 323}, 225 (1989).
}

\bibitem{FE5}
{
 F. Embacher,
 {\it preprint} gr-qc/9504040.
}


\bibitem{555}
{
 L. P. Grishchuk and L. V. Rozhansky,
 {\it Phys. Lett. B} {\bf 208}, 369 (1988);
 {\bf 234}, 9 (1990).
 A. Lukas,
 {\it preprint} gr-qc/9409012.
}

\bibitem{666}
{
 A. Vilenkin,
 {\it Phys. Rev. D} {\bf 30}, 509 (1984);
 {\bf 33}, 3560 (1986).
 A. D. Linde,
 {\it Nuovo Cimento} {\bf 39}, 401 (1984).
}


\end{thebibliography}
\end{document}